\documentclass[preprintnumbers,amsmath,amssymb,twocolumn]{revtex4}
\usepackage[dvips,colorlinks=true,citecolor=red,filecolor=green,linkcolor=blue,pdfnewwindow=true]{hyperref}
\usepackage{graphicx}
\usepackage{makeidx}
\usepackage{bm}
\usepackage[title]{appendix}

\begin{document}
\title{Temperature driven metal-insulator transition in thin films}
\author{R.K. Brojen Singh}
\email{brojen@jnu.ac.in}
\affiliation{School of Computational and Integrative Sciences, Jawaharlal Nehru University, New Delhi-110067, India.}

\begin{abstract}
We present the results of electron delocalization in thin films of finite thickness at finite temparature induced by thickness of the film and temparature. The two dimensional results show temparature induced metal-insulator transition (MIT) obtaining both insulating and metallic solutions. The localization length in insulating regime and zero frequency conductivity in metallic regime are found to be as a function of temparature and disorder parameter. Similarly, in thin films of finite thickness, delocalization of electronic states takes place induced by temparature as well as film thickness. Further, we could able to get critical temparature $T_c$ for fixed thickness and critical thickness, $b_c$ for fixed temparature. In metallic regime, conductivity is found to be as a function of film thickness, temparature and disorder.
\end{abstract}


\maketitle

\section{Introduction}
The scaling theory of localization proposed that all single particle quantum states of an electronic band in two dimensional noninteracting electrons solid are localized for any disorder however week but in three dimensional system there is a critical disorder above which all the states are localized, and below which some states are delocalized \cite{aalr}. The temparature dependence of Drude conductivity in two dimensional system which comes from phase breaking time was studied in 2D-Anderson model of noninterating electron gas \cite{aalr,gdd} and 2D metal-insulator transition scenario in the charged trap model \cite{am}. Numerically at finite temparature it was found logarithmic dependence of temparature of 2D conductivity and $T^{p/2}$ dependence to the conductivity correction in 3D, where $p$ is phase breaking exponent \cite{mk}. Signature of weak localization correction to resistivity in thin metallic films at low temparatures was seen \cite{berg}. Experimentally conductance fluctuation in quasi-one-dimensional inversion layers in Si-MOSFETs was also observed at low temparatures \cite{fhw}. 

However, the electron transport mechanism in thin films in the presence of temparature is still unclear specially in weak disorder regime. The quantum interference effects in these systems play an important role in the study of metal-insulator transition. But a systematic study showing clear understanding of electron localization in these systems are still remain unfinished. We address these questions and examine transport behavior of noninteracting electrons in weakly disordered quasi-two dimensional layers or thin films induced by temparature. 

Our work is planned as in the following. In section 2 we briefly review self consistent theory of localization in disordered thin films. The systematic derivation of localization length as a function of various parameters, such as film thickness, disorder parameter and temparature is done in this section. Section 3 provides the two dimensional solutions both in insulating and metallic regimes induced by temparature are presented. The thin film solution both in insulating and metallic regime are discussed in section 4. In section 5 we draw some conclusion based on our results obtained in the respective sections.

\section{Self-consistent theory in thin films}
The scaling ideas at the microscopic level, have been supported by the self-consistent theory of localization due to Vollhardt and Wolfle \cite{vw1,vw2} $(VW)$. This theory is very convenient technique to study metal-insulator transition specially to deal with weak disorders. In this theory one calculates the density response function, which is related to particle-hole pair propagator. The density fluctuations propagate diffusively due to particle number conservation, with a diffusion constant $D(\vec q,\omega )$, which gets strong size and dimension dependent quantum corrections from the vertex in the particle-particle channel. These corrections are account for the enhanced interference between the time reversed paths and also have the same diffusive character when time-reversal invariance is present. VW theory was extended by Yoshioka, Ono and Fukuyama \cite{yof} to situations when the time reversal invariance is absent. It was argued that when the time-reversal invariance is not present the particle-hole and particle-particle channels are not related and one needs two diffusion constants. In this case self-consistent relation of VW gets replaced by a set of two relations between the diffusion constants of the two channels. Whereas in the absence of time-reversal invariance the two diffusion constants becomes equal and $VW$ theory explains the situation. The theory was then extended to thin film of finite thickness in the absence of any field \cite{bk1} and in the presence of perpendicular magnetic field \cite{bk2}. 

The equation for the frequency dependent diffusion constant, $D(\omega)$ incorporating quantum diffusion due to backscattering, can be derived self consistently \cite{vw1,vw2} to arrive at,
\begin{eqnarray}
\frac{D(\omega)}{D_0}=1-\lambda dk^{2-d}_F\int^{1/l}_{0}dq\frac{q^{d-1}}{-\frac{i\omega}{D(\omega)}+q^2}
\label{dif}
\end{eqnarray}
where, $\lambda=(2\pi E_F\tau)^{-1}$ is disorder strength in terms of Fermi energy $E_F$ and the collision time $\tau$ and $k_F$ is Fermi wave number. $D_0$ is bare diffusion constant. The integration is restricted to momenta smaller than the inverse of the mean free path, $1/l$. This equation for $D(\omega)$ gives rise metallic solution when $D(\omega)\rightarrow D_0$ as $\omega\rightarrow 0$, is a positive number, whereas insulating solution if $-D(\omega)/i\omega=\xi^2(\omega)\rightarrow\xi^2(0)$, a real positive number.

The backscattering corrections in weak scattering regime for restricted geometries can be calculated systematically using path integral method \cite{cs}. This method assumes that the classical paths in the presence of randomly placed impurities can be taken as random walks \cite{berg, kl, kh}. Then the quantum corrections are related to the probability of return of the walk to its origin, which is easily obtained by solving the diffusion equation in the desired geometry \cite{cs}. If we denote the coordinate of the film be $(\vec r,z)$, where $\vec r$ is vector along the plane and $z$ is the coordinate along the thickness, one is able to calculate the probability of return, $P(\vec r,z,t)$ to the point $(\vec r,z)$ after time $t$, by solving the diffusion equation \cite{cs}. The boundary condition to be applied is that the current normal to the surfaces of the film vanishes. The solution for $P(\vec r,z,t)$ is given by, 
\begin{eqnarray}
P(\vec r,z,t)&=&\frac{2}{Ab}\sum_{n=0}^{\infty}\sum_{q}cos^2(\frac{\pi nz}{b}) \nonumber \\
&&\times exp\left[-D_0\left(q^2+\frac{\pi^2n^2}{b^2}\right)t\right]
\label{prob}
\end{eqnarray}
where $A$ and $b$ are area and thickness of the film respectively. $\vec q$ is two dimensional wave vector. For self-consistency, $D_0$ is replaced by $D(\omega)$ and incorporating this result one obtains the following equation,
\begin{eqnarray}
\label{s7}
\frac{D(\omega)}{D_0}=1-2\lambda\int_{0}^{1/l}dqq\sum_{n=0}^{\infty}\frac{1}{q^2
+\left(\frac{n\pi}{b}\right)^2-\frac{i\omega}{D(\omega)}}
\end{eqnarray}
For finite size of the system, the lower limit of the integration in equation (\ref{s7}) is replaced by $1/L$. If we define, $-D(\omega)/i\omega =\xi^2(\omega)$ and using the following summation formula,
\begin{eqnarray}
\sum_{n=0}^{\infty}\frac{1}{c^2+n^2\pi^2}=\frac{1}{2}\left(\frac{coth(c)}{c}+\frac{1}{c^2}\right)
\end{eqnarray}
and by performing wave vector integration, one can easily able to get integral equation for $\xi(\omega)$. The result is given by,
\begin{eqnarray}
\label{s18}
\frac{1}{\lambda}=\log\left[\sqrt{\frac{1+\tilde\xi(\omega)^{-2}}
{\tilde L^{-2}+\tilde\xi(\omega)^{-2}}}
\times\frac{\sinh\left(\tilde b\sqrt{1+\tilde\xi(\omega)^{-2}}\right)}{\sinh\left(\tilde b
\sqrt{\tilde L^{-2}+\tilde\xi(\omega)^{-2}}\right)} \right]
\end{eqnarray}
where, $\tilde\xi(\omega)=\xi(\omega)/l$, $\tilde b=b/l$ and $\tilde L=L/l$ which are dimensionless parameters. Here $l$ is the elastic mean free path. This situation is physically relevent at finite temparatures. Due to inelastic scattering, the electron looses its phase coherence over a distance $L_\phi$, the inelastic phase coherence length. Quantum interference effects are believed to be observable only at low temparature \cite{kram}. The coherence regime can be defined by the condition $\tilde L_\phi\ge\tilde L$. So we replace $\tilde L$ in this equation by $L_\phi$. This $L_\phi$ depends upon temparature which can be taken as $L_\phi\approx L_0(T/T_0)^{-p}$, where $p$ is a parameter depending on scattering mechanism, dimensionality, etc. $L_0$ and $T_0$ are constants having the dimensions of length and temparature respectively. To study cross-over from two to three dimensions we need to scale the disorder parameter, $\lambda$ such that $\lambda^{-1}=\hbar N_F(2)D_0$ and $4/(3\pi l\lambda^{2})=\hbar N_F(3)D_0$ where, $N_F(2)$ and $N_F(3)$ are density of states at Fermi level at two and three dimensions respectively \cite{bk1,bk2,rkbs}. In this situation, we study Metal-Insulator transition induced by temparature in two-dimensional as well as in thin films respectively.

\section{\textbf{MIT in 2D electron system}}

\begin{figure}[htbp]
\centering
\includegraphics[width=3.3in]{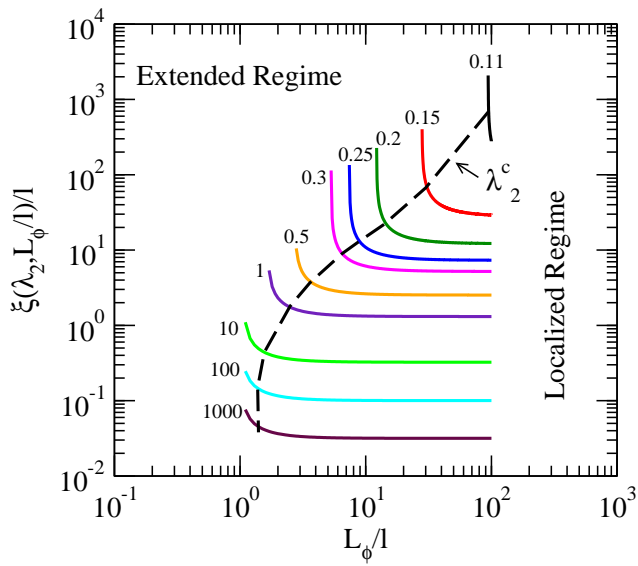}
\caption{Plots of localization length $\xi$ in two dimensional electrons system as a function of $L_\phi$ for two different values of $\lambda_2$.}
\label{ssc8}
\end{figure}

We obtain two dimensional insulating solution as $\omega\rightarrow 0$ and by taking the limit $\tilde b\rightarrow 0$ in equation (\ref{s18}) and replacing $\tilde L$ by $\tilde L_\phi$. In this limit one can approximate sinh function as exponential and $\lambda\rightarrow\lambda_2$. Then solving for $\tilde\xi(0)=\tilde\xi$,  we get 
\begin{eqnarray}
\label{ss20}
\tilde\xi(\lambda_2,\tilde L_\phi)=\tilde L_\phi\sqrt{\frac{e^{1/\lambda_2}-1}{\tilde L_\phi^{2}
-e^{1/\lambda_2}}}
\end{eqnarray}
The numerator inside the square root of this equation is always positive number for any values of $\lambda_2$. The solution of $\tilde\xi$ as a function of $\tilde L_\phi$ for various values of $\lambda_2$ is shown in Fig.\ref{ssc8}. The plots show that as $\tilde L_\phi$ decreases i.e. temparature, $T$ increases $\tilde\xi$ remains almost stationary for some range of $\tilde L_\phi$ and then $\tilde\xi$ increases monotonically (showing divergence of $\tilde\xi$ at different $L_\phi$ values for different $\lambda_2$s) as $T$ increases. This provides the signature of metal insulator transition in 2D system induced by $T$ for different values of $\lambda_2$. The critical values $\tilde L_\phi^c$ and $\lambda_2^c$ can be calculated at $\tilde\xi\rightarrow\infty$ from equation (\ref{ss20}), and is given by, $\frac{1}{\lambda_2^c}=2log(\tilde L^c_\phi)$. The dotted line is the approximate critical line which separate extended and localized regimes in 2D electron system. Then from equation $\ref{ss20}$ we get real positive solution of $\tilde\xi$ only when the denominator is greater than one. So we get insulating phase as long as,
\begin{eqnarray}
\label{sss21}
\frac{L_\phi^2}{l^2}> e^{1/\lambda_2}
\end{eqnarray}
With increasing temparature T, $\tilde L_\phi$ decreases and at a certain critical temparature $T_c(\lambda_2)$, this inequality is no longer satisfied and $\tilde\xi$ does not have solution. It happens when the inequality (\ref{sss21}) becomes equality giving 
$T_c(\lambda_2)$,
\begin{eqnarray}
\label{sss22}
T_c(\lambda_2)=T_0\left(\frac{L_0}{l}\right)^{2/p}e^{-1/(p\lambda_2)}
=Ce^{-1/(p\lambda_2)}
\end{eqnarray}
Where, $C=T_0\left(\frac{L_0}{l}\right)^{2/p}$ is a constant. When $T>T_c$, the equation (\ref{ss20}) does not have solutions. In another words, metallic solution in two-dimensional case is obtained when 
\begin{eqnarray}
\label{sss23}
\frac{L_\phi^2}{l^2}< e^{1/\lambda_2}
\end{eqnarray}
In metallic phase we solve for real and positive value of D(0) in the limit $\omega\rightarrow 0$. Similarly for a fixed temparature one can also obtain these two insulating and metallic phases induced by $\lambda$ which can be seen from equation $(\ref{ss20})$. In this situation taking $\tilde\xi\rightarrow\infty$ we obtain critical disorder, $\lambda_c$ from equation $(\ref{ss20})$ as, $\lambda_c=1/(2log(\tilde L_\phi))$. Then we calculate two dimensional zero frequency conductivity, $\sigma_2$ using Einstein's relation, $\sigma_2(\omega)=e^2N_F(2)D(\omega)$ as $\omega\rightarrow 0$ and substituting the value of $\lambda_c$. The result is
\begin{eqnarray}
\label{ss22}
\sigma_2(\lambda_2,\tilde L_\phi)=\sigma_{02}\left[1-2\lambda_2\log(\tilde L_\phi)\right]
=\sigma_{02}\left[1-\frac{\lambda_2}{\lambda_2^c}\right]
\end{eqnarray} 
where, $\sigma_{02}=e^2N_F(2)D_0$ is zero frequency Drude's conductivity in two dimensional system. Thus obviously we get a metal to insulator transition induced by temparature as well as disorder in two-dimensional system.

\section{MIT in Thin film}

\begin{figure}[htbp]
\centering
\includegraphics[width=3.2in]{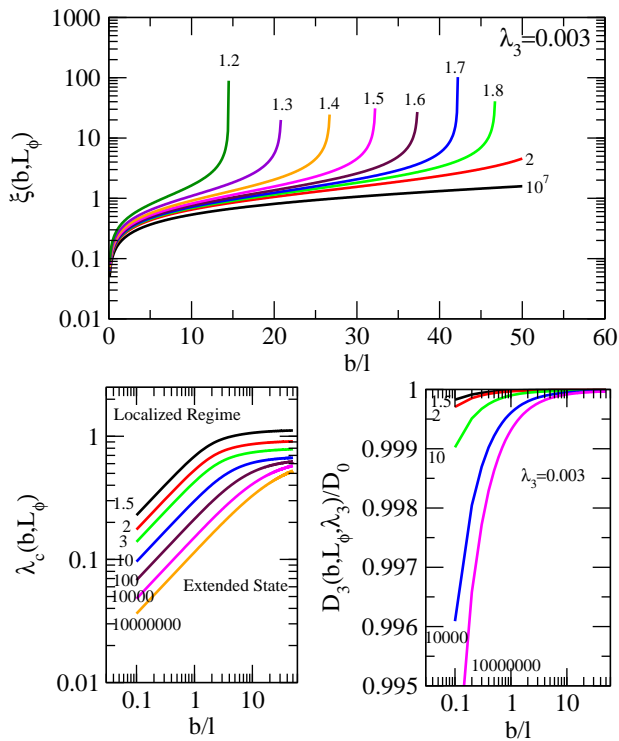}
\caption{Plots of localization length $\xi$ as a function of $b/l$ for two
different values of $L_\phi$ for fixed value of $\lambda_3=0.43$.}
\label{sc8}
\end{figure}

The insulating solution of thin film can be obtained by scaling $\lambda$ given by, $4/(3\pi l\lambda_3^{2})=\hbar N_F(3)D_0$ and using the relation $(\ref{s18})$ as $\omega\rightarrow 0$ by
\begin{eqnarray}
\label{ss18}
\frac{1}{\lambda_3}&=&\frac{3\pi}{4\tilde b}\log\left[\sqrt{\frac{1+\tilde\xi^{-2}}
{\tilde L_\phi^{-2}+\tilde\xi^{-2}}}
\times\frac{\sinh\left(\tilde b\sqrt{1+\tilde\xi^{-2}}\right)}{\sinh\left(\tilde b
\sqrt{\tilde L_\phi^{-2}+\tilde\xi^{-2}}\right)} \right]\nonumber\\
&=&\frac{3\pi}{4\tilde b}\log\left(\frac{1+\tilde\xi^{-2}}{\tilde L_\phi^{-2}+\tilde\xi^{-2}}\right),~~~~~~~~~~\tilde b\langle\langle 1
\end{eqnarray}
The behaviour of $\tilde\xi$ with respect to $\tilde b$ is shown in the Fig.(\ref{sc8}). In this figure we found that, for a fixed value of $\lambda_3$, $\tilde\xi$ diverges at different values of $\tilde b$ for various values of $\tilde L_\phi$ ($\langle 10^7$) as shown in Fig. \ref{sc8}. Whereas for large values of $\tilde L_\phi$ ($\rangle 10^7$), $\tilde\xi$ saturates with $\tilde b$ showing the signature of existance of insulating phase driven by $T$. Further, the equation ($\ref{ss18}$) indicates that $\tilde\xi$ always does not have solution for any values of $\tilde b$, $\lambda_3$ and $\tilde L_\phi$. This shows the possibility of metal insulator transition in thin films induced by $T$. 

The critical disorder, which can be defined as the disorder at which phase transition takes place, can be obtained from equation (\ref{ss18}) by taking $\tilde\xi\rightarrow\infty$ and the critical disorder $\lambda_c$ is given by,
\begin{eqnarray}
\label{s20}
\frac{1}{\lambda_3^c(\tilde b,L_\phi)^2}&=&\frac{3\pi}{4\tilde b}\log\left(\tilde L_\phi\times\frac{\sinh(\tilde b)}{\sinh(\tilde b/\tilde L_\phi)}\right)\nonumber\\
&=&\frac{3\pi}{2\tilde b}\log\left(\tilde L_\phi\right),~~~~~~~~~~\tilde b\langle\langle 1
\end{eqnarray}
The critical disorder is found to be as a function of $\tilde b$ and $\tilde L_\phi$. So for a finite temparature, one can obtain thickness induced delocalization of states in thin films. The behaviour of $\lambda_3^c$ as a function of $\tilde b$ for different values of $L_\phi(T)$ is shown in Fig. \ref{sc8} (lower left panel) separating localized and extended states in the phase diagram. From this figure one can able to see that as temparature increases, the critical disorder increases and saturates. Since we have a phase transition in this case, it is possible to obtain zero frequency diffusion constant in extended regime. We can calculate it from equation (\ref{s7}) first by taking the limit $\omega\rightarrow 0$ and then doing the summation and integration respectively. Then using equation ($\ref{s20}$) we get the following result,
\begin{eqnarray}
\label{s21}
D_3(\lambda_3,\tilde b,\tilde L_\phi)&=&D_0\left[1-\frac{3\pi}{4\tilde b}\lambda_3^2
\log\left(\tilde L_\phi\times\frac{\sinh(\tilde b)}{\sinh(\tilde b/\tilde L_\phi)}
\right)\right]\nonumber \\
&=&D_0\left[1-\left(\frac{\lambda_3}{\lambda_c(\tilde b,\tilde L_\phi)}\right)^2\right]
\end{eqnarray}
where, the finite thickness zero frequency diffusion constant, $D_3$ is found to be as a function of $\tilde b$, $\lambda_3$ and $\tilde L_\phi$. We can also get metallic solution only when $\lambda_3<\lambda_c$. Fig. \ref{sc8} (lower right panel) shows the solution of $D_3$ with respect to $\tilde b$ for different values of $\tilde L_\phi$ and fixed $\lambda_3$. The curves monotonically increases as $\tilde b$ increases and then starts saturating to some value as $\tilde b$ increases. Now by straightforward using equation (\ref{s21}) and the Einstein's relation of conductivity, we can obtain zero frequency conductivity $\sigma_3$ as in the following,
\begin{eqnarray}
\label{s22}
\sigma_3(\lambda_3,\tilde b,\tilde L_\phi)=-\sigma_{03}\left[1-\left(\frac{\lambda_3}{\lambda_c(\tilde b, \tilde L_\phi)}\right)^2\right]
\end{eqnarray}
where, $\sigma_{03}=e^2N_F(3)D_0$ is zero frequency Drude's conductivity in three dimensional system. So clearly in the finite system size, there is a Metal-Insulator transition driven by temparature, thickness of the film and disorder.

\section{Conclusion}

We have extended self-consistent theory of localization due to Vollhardt and W\"olfle to thin films incorporating temperature $T$ and studied the role of temparature in metal-insulator transition in 2D and thin layered films. We analyzed the phase transition when phase relaxation length $\tilde L_\phi$ is finite.

In two dimensions, we found localization length in insulating regime as a function of disorder as well as temparature. We obtain metal-insulator transition in 2D induced by disorder as well as temparature. Further, we could able to get critical disorder as a function of temparature in 2D system. We also calculated zero frequency diffusion constant and conductivity in metallic regime. In the case of thin films, we found a critical disorder which depends on thickness of the film and temparature. In this case we obtained a transition from insulator to metal driven by thickness and temparature. In insulating regime, the localization lengths increases rapidly as a function of thickness as temperature $T$ increases for a fixed value of disorder. But for small $T$, the localization length saturates to some value as we increase thickness. We calculated the zero frequency conductivity for the disorder smaller than the critical value (metallic regime). We claim that for non-zero temparature, there is possibility of insulator to metal transition induced by thickness of the film as well as temparature.

\end{document}